\documentclass{appolb}
\usepackage{graphicx}
\usepackage{stmaryrd}
\usepackage{comment}
\usepackage{color}
\usepackage{amsmath}
\usepackage{amssymb}
\usepackage{amsthm}
\usepackage{mathrsfs}
\usepackage{graphicx}
\usepackage{marvosym}
\usepackage{amsmath,bm}

\begin{document}
\title{Spin alignment from turbulent color fields
\thanks{Presented at Quark Matter 2022}%
}
\author{Berndt M\"uller$^1$, Di-Lun Yang$^2$
\address{$^1$Department of Physics, Duke University, Durham, North Carolina 27708, USA.\\
	$^2$Institute of Physics, Academia Sinica, Taipei 11529, Taiwan.
}
}
\maketitle
\begin{abstract}
We study the important, yet widely overlooked, role of gluons for spin transport with a connection to local parity violation in quark gluon plasmas. We extend the newly developed quantum kinetic theory for relativistic fermions to the case coupled with non-Abelian chromo-electromagnetic fields and employ this formalism to investigate the spin polarization of quarks under dynamically generated color fields in near-equilibrium quark gluon plasmas. It is found that the spin polarization could be induced by parity-odd correlators of color fields, which may dominate over collisional effects at weak coupling. Our result provides with a possible explanation for the spin alignment of vector mesons measured in high-energy nuclear collisions and alludes to the connection with local parity violation.
\end{abstract}
  
\section{Introduction}
Recent observations for the spin alignments of vector mesons in RHIC \cite{STAR:2022fan} and LHC \cite{ALICE:2019aid} have raised a new puzzle for theoretical explanations. The measured longitudinal (00)
element of the spin density matrix of the vector meson is related to the polarization of a quark and an anti-quark through $\rho_{00}=(1-\mathcal{P}^{i}_{q}\mathcal{P}^{i}_{\bar{q}})/(3+\mathcal{P}^{i}_{q}\mathcal{P}^{i}_{\bar{q}})$ based on the coalescence model \cite{Liang:2004xn} (see \cite{Sheng:2022ffb} for a generic relation), where $i$ corresponds to the spin quantization axis. The large deviation of $\rho_{00}$ from $1/3$ implies relatively strong spin polarization of quarks and anti-quarks that cannot be simply explained by spin polarization from vorticity at least for high-energy collisions. Moreover, the sign of $\rho_{00}-1/3$ also depends on collision energies, species of vector mesons, and their transverse momenta. It is hence rather challenging, and perhaps unpractical, to find a single effect as the universal explanation. 
See, e.g., \cite{Sheng:2019kmk,Sheng:2020ghv,Xia:2020tyd,Sheng:2022wsy} for several proposed mechanisms.  

On the other hand, it is theoretically important to understand the dynamical spin polarization of a quark traversing the quark gluon plasma (QGP). While the spin polarization induced by collisions between quarks in connection to spin-orbit interaction has been intensively studied especially in the framework of the quantum kinetic theory (QKT) (see \cite{Hidaka:2022dmn} for a review), how gluons can indirectly affect the spin transport of quarks remains unclear. Except for possible collisional effects led by polarized gluons as suggested by the QKT of photons in QED \cite{Hattori:2020gqh,Lin:2021mvw}, we will review how the low-energy gluons characterized by turbulent color fields in the weakly-coupled QGP could trigger spin polarization of quarks and affect spin alignment of vector mesons \cite{Muller:2021hpe,Yang:2021fea}.   
\section{QKT with color fields}
In the early stage of relativistic heavy ion collisions, strong color fields (or chromo-electromagnetic fields) could emerge in the glamsa phase \cite{Gelis:2010nm}, which however decay with time. Nevertheless, the Weibel-type plasma instability due to anisotropy in the QGP phase may dynamically generate the color fields \cite{Mrowczynski:1988dz,Mrowczynski:1993qm,Romatschke:2003ms}. It was proposed in \cite{Asakawa:2006tc,Asakawa:2006jn} such turbulent color fields could result in anomalous dissipative transport that may dominate over the collisional effects in weakly coupled QGP. The dynamics of an onshell quark is then described by the modified Boltzmann equations,
\begin{eqnarray}
	\label{SKE_singlet_simplify}
	p\cdot\partial f^{s}_V(p,X)-\partial_{p}^{\kappa}\mathscr{D}_{\kappa}[f_V^{s}]=\mathcal{C}[f_{V}^{s}],
\end{eqnarray}
where 
\begin{eqnarray}
	\mathscr{D}_{\kappa}[O]=\frac{g^2}{6}\int^{p}_{k,X'}p^{\lambda}p^{\rho}\langle F^a_{\kappa\lambda}(X)
	F^a_{\alpha\rho}(X')\rangle \partial_{p}^{\alpha}O(p,X')
\end{eqnarray}
and $F^a_{\kappa\lambda}$ is the field strength of color fields with the superscript $a$ denoting the color index. 
Here $f^{s}_V(p,X)$ represents the color-singlet (vector-charge) distribution function of a quark and $\mathcal{C}[f_{V}^{s}]$ denotes the collision kernel.\footnote{The quark distribution function is generically a matrix in color space, which can be decomposed into the color singlet and color-octet components governed by coupled kinetic equations \cite{Heinz:1983nx,Elze:1986qd}. At weak coupling, one could rewrite the color-octet component in terms of the color-singlet one by perturbatively solving the color-octet kinetic equation.} We also introduced the abbreviation
\begin{equation}
	\int^{p}_{k,X'}\equiv \int\frac{d^4k d^4X'}{(2\pi)^4}e^{ik\cdot(X'-X)}\left(\pi\delta(p\cdot k)+i\frac{{\cal P}}{p\cdot k}\right)
\end{equation}
with ${\cal P}$ denoting the principal value. The effective diffusion term $\partial_{p}^{\kappa}\mathscr{D}_{\kappa}[f_V^{s}]$ with the color-field correlator stems from the correlation of two non-local (color) Lorentz forces acting on the quark. Such background color fields  originate from soft gluons emitted by stochastic sources in the QGP, whereas the scattering with hard partons is given by $\mathcal{C}[f_{V}^{s}]$. Assuming the color-field correlator is greater than $O(g^2)$ due to the strong plasma instability, the diffusion term in (\ref{SKE_singlet_simplify}) may dominate over $\mathcal{C}[f_{V}^{s}]\sim O(g^4)$ at weak coupling and the kinetic equation reduces to a Fokker-Plank equation.  

By analogy with \cite{Asakawa:2006tc,Asakawa:2006jn}, it is anticipated that the spin transport may be affected by the color-field correlator, whereas the origin should be quantum corrections beyond just the classical Lorentz forces. In order to track the spin transport of a massive quark, the so-called axial-kinetic theory \cite{Hattori:2019ahi,Yang:2020hri} is adopted and generalized to include background color fields \cite{Yang:2021fea}. One could also construct the similar kinetic theory for massless quarks \cite{Luo:2021uog} based on the chiral kinetic theory \cite{Son:2012wh,Stephanov:2012ki,Chen:2012ca,Hidaka:2016yjf,Hidaka:2017auj}. For massive quarks, except for (\ref{SKE_singlet_simplify}) delineating (vector) charge, energy, and momentum transport, the dynamical spin evolution in phase space is captured by an axial-vector kinetic equation \cite{Muller:2021hpe,Yang:2021fea},
\begin{eqnarray}\label{AKE_singlet_simplify}
	0&=&p\cdot\partial\tilde{a}^{s\mu}(p,X)-\partial_{p}^{\kappa}\mathscr{D}_{\kappa}[\tilde{a}^{s\mu}]
	+\partial_{p}^{\kappa}\big(\mathscr{A}^{\mu}_{\kappa}[f^{\rm s}_{V}]\big),
\end{eqnarray} 
where 
\begin{eqnarray}\nonumber
	\mathscr{A}^{\mu}_{\kappa}[O]&=&\frac{g^2}{6}
	\epsilon^{\mu\nu\rho\sigma}\int^{p}_{k,X'}p^{\lambda}p_{\rho}\Big(\partial_{X'\sigma}\langle F^a_{\kappa\lambda}(X)F^a_{\alpha\nu}(X')\rangle
	\\
	&&+\partial_{X\sigma}\langle F^a_{\kappa\nu}(X)F^a_{\alpha\lambda}(X')\rangle\Big)\partial^{\alpha}_{p}O(p,X').
\end{eqnarray}
Here the axial vector $\tilde{a}^{s\mu}(p,X)$ is an effective (color-singlet) spin four vector contributing to spin polarization and axial-charge currents, which satisfies $p\cdot \tilde{a}^{s}=0$ under the onshell condition $p_0=\epsilon_{\bm p}\equiv \sqrt{|\bm p|^2+m^2}$. We omitted the collision term suppressed at weak coupling by the aforementioned assumption. Except for the diffusion term $\partial_{p}^{\kappa}\mathscr{D}_{\kappa}[\tilde{a}^{s\mu}]$, $\partial_{p}^{\kappa}\big(\mathscr{A}^{\mu}_{\kappa}[f^{\rm s}_{V}]\big)$ serves as a source term that dynamically induces nonzero $\tilde{a}^{s\mu}$ from $f_V^s$, where the color-field correlators therein come from the correlation between a Lorentz force and a quantum correction such as the term responsible for a spin-Hall effect.  
\section{Source terms and spin polarization}
Given $f^s_V$ and $\tilde{a}^{s}_{\mu}$ obtained from (\ref{SKE_singlet_simplify}) and (\ref{AKE_singlet_simplify}), one can evaluate the spectrum of spin polarization via \cite{Muller:2021hpe,Yang:2021fea}
\begin{equation}\label{Spin_CooperFrye}
	\mathcal{P}^{\mu}({\bf p})=\frac{\int d\Sigma\cdot p\mathcal{J}_{5}^{\mu}(p,X)}{2m\int d\Sigma\cdot\mathcal{N}(p,X)}\bigg|_{p_0=\epsilon_{\bm p}}
\end{equation}
by integrating over a freeze-out hypersurface $\Sigma_{\mu}$,
where 
\begin{eqnarray}
	\mathcal{N}^{\mu}(p,X)&=& p^{\mu}f^s_{V},
	\\\label{eq:J5mu_density}
	\mathcal{J}_5^{\mu}(p,X)&=&\big(\tilde{a}^{ s\mu}+g^2(\mathcal{A}^{\mu}_{Q}+\mathcal{A}^{\mu}_{S})/6\big).
\end{eqnarray}
The coefficients $\mathcal{A}^{\mu}_{Q}$ and $\mathcal{A}^{\mu}_{S}$ explicitly read
\begin{equation}\label{AQmu_origin}
	\mathcal{A}^{\mu}_{Q}=\frac{\partial_{p\kappa}}{2}\int^{p}_{k,X'}p^{\beta}\langle \tilde{F}^{a\mu\kappa}(X)F^a_{\alpha\beta}(X')\rangle\partial_{p}^{\alpha}f^{\rm s}_V(p,X')
\end{equation}
and
\begin{eqnarray}
	\mathcal{A}^{\mu}_{S}=-\frac{p_0}{2}\left(\partial_{pj}-\frac{p_{j}}{p_0}\partial_{p0}\right)\bigg[\int^{p}_{k,X'}\frac{p^{\beta}}{p_0}
	\langle \tilde{F}^{a\mu j}(X)F^a_{\alpha\beta}(X')\rangle\partial_{p}^{\alpha}f^{\rm s}_V(p,X')\bigg]
\end{eqnarray}
with $j=1,2,3$ and $\tilde{F}^{a\mu\nu}\equiv \epsilon^{\mu\nu\alpha\beta}F^a_{\alpha\beta}/2$. Notably, unlike $\mathcal{A}^{\mu}_{Q}$, which could further yield an axial charge current, $\mathcal{A}^{\mu}_{S}$ comes from a surface term, which modifies the spectrum without affecting the current. Here $\mathcal{A}^{\mu}_{Q}+\mathcal{A}^{\mu}_{S}$ engenders non-dynamical spin polarization even when $\tilde{a}^{s\mu}=0$. We will further explore how the spin polarization induced by turbulent color fields could affect the spin alignment with an emphasis on the case in LHC energies.  

In practice, the realistic color-field correlators may be acquired from real-time
simulations for prescribed initial conditions. We will instead make physical assumptions to derive an analytic expression. Assuming local spacetime translational invariance of the QGP, we propose 
\begin{equation}
	\langle F^a_{\kappa\lambda}(X)F^a_{\alpha\rho}(X')\rangle = \langle F^a_{\kappa\lambda}F^a_{\alpha\rho}\rangle e^{-(t-t')^2/\tau_c^2} ,
\end{equation} 
where $\tau_c$ denotes the correlation time. We may further consider the case when $f^s_{V}$ reach thermal equilibrium and thus take $f^{\rm s}_{V}(p)=f_{\rm eq}(p\cdot u)\equiv (e^{(p\cdot u-\mu)/T}+1)^{-1}$, where $T$, $\mu$, and $u^{\mu}$ correspond to the temperature, chemical potential, and fluid four velocity, respectively. The local fluctuations could further yield anomalous dissipative transport processes \cite{Asakawa:2006tc}, while we will neglect the gradient terms of hydrodynamic variable for simplicity. Note that vorticity is also expected to be small in high-energy collisions. We next introduce the (chromo-) electric and magnetic fields in the QGP rest frame, $E^{a\mu}=F^{a\mu\nu}u_{\nu}$ and $B^{a\mu}=\epsilon^{\mu\nu\alpha\beta}u_{\nu}F^a_{\alpha\beta}/2$. Furthermore, the hierarchy $|\langle B^{a}_{\mu}B^{a}_{\nu}\rangle| \gg |\langle E^{a}_{\mu}B^{a}_{\nu}\rangle| \gg |\langle E^{a}_{\mu}E^{a}_{\nu}\rangle|$ is postulated, stemming from the screening of the chromo-electric field as opposed to the chromo-magnetic field albeit in the static case \cite{Weldon:1982aq} and amplification of the latter from plasma instability in anisotropic QGP \cite{Mrowczynski:1993qm}. The symmetric condition $\langle E^a_{\mu}B^{a}_{\nu}\rangle=\langle E^a_{\nu}B^{a}_{\mu}\rangle$ is also imposed for simplicity.

Based on the approximations above and working in the fluid rest frame, the solution of (\ref{AKE_singlet_simplify}) yields \cite{Muller:2021hpe,Yang:2021fea}
\begin{eqnarray}\label{amu_from_source}
	\tilde{a}^{s\mu}(t,p)=-\frac{\hbar(t-t_0)}{12p_0^2}
	\big(\partial_{p0}f_{\rm eq}(p_0)\big)\big(\langle B^{a\mu}E^{a\nu}\rangle p_{\nu}-\langle B^a\cdot E^a\rangle p^{\mu}_{\perp}\big),
\end{eqnarray}
where the secular term stems from translational invariance in time.
Additionally, the onshell non-dynamical source term in $\mathcal{J}_5^{\mu}(p,X)$ becomes \cite{Muller:2021hpe,Yang:2021fea}
\begin{eqnarray}\nonumber\label{AQmu}
	(\mathcal{A}^{\mu}_{Q})_{\rm eq}+(\mathcal{A}^{\mu}_{S})_{\rm eq}&\approx&\frac{\sqrt{\pi}\tau_c}{4\epsilon_{\bm p}^3}\Big[(p^{\alpha}p^{\beta}\langle E^a_{\alpha}B^a_{\beta}\rangle u^{\mu}
	-\langle B^{a\mu}E^{a\nu}\rangle
	\\
	&&\times \epsilon_{\bm p}(\epsilon_{\bm p}^2\partial_{p\nu}-p_{\nu})\Big]
	\partial_{\epsilon_{\bm p}}f_{\rm eq}(\epsilon_{\bm p}).
\end{eqnarray}
It is found that the spin polarization of a quark, from (\ref{Spin_CooperFrye}), could be engendered by parity-odd correlators of color fields. Because $\mathcal{P}^{i}_{q}=\mathcal{P}^{i}_{\bar{q}}$ and $|\mathcal{P}^i_{q/\bar{q}}|$ is larger for light quarks, we expect $\rho_{00}(K^{*0})< \rho_{00}(\phi)<1/3$ from \cite{Liang:2004xn}. This qualitative behavior is consistent with the ALICE measurements \cite{ALICE:2019aid}. Note that here the spin polarization in a quadratic form of color fields in QCD is intrinsically different from the polarization linear to an $U(1)$ electromagnetic field in QED.    
\section{Discussions and outlook}
Notably, the presence of parity-odd color-field correlators requires local-parity violation of the QGP. Since such correlators fluctuate event by event, they only contribute to spin alignment without affecting the spin polarization of $\Lambda$ hyperons. The finding reveals a possible relation between spin alignment and local parity violation (see also \cite{Gao:2021rom}), even for out-off equilibrium conditions, in QCD plasmas as a long-sought problem.  

Nonetheless, the result is obtained under several assumptions and approximations, among which the input of more realistic color-field correlators will be essential for future study. On the other hand, unlike the non-dynamical source term dictated by color fields at freeze-out, the dynamical spin polarization from $\tilde{a}^{s\mu}$ has the memory effect as manifested by the secular solution, which implies that $\tilde{a}^{s\mu}$ may be influenced by even strong color fields in the glasma phase and the accumulated spin polarization may play a more dominant role than the non-dynamical contribution. It is tempting to further investigate such an effect on dynamical spin polarization, which could exist even in the absence of turbulent color fields in the QGP phase.    

\textit{Acknowledgments}.
This work is supported by the U. S. Department of Energy under Grant No. DE-FG02-05ER41367
and Ministry of Science and Technology, Taiwan under Grant No. MOST 110-2112-M-00l-070-MY3.

\bibliographystyle{h-physrev}
\bibliography{Dl_Yang_QM22.bbl}

\end{document}